\documentclass[aps,prl,amsmath,amssymb,twocolumn,showpacs,byrevtex,superscriptaddress]{revtex4}
\usepackage{graphicx}
\begin{document}
\title{
Diffusivity and configurational entropy maxima in short range attractive colloids
}

\author{L.~Angelani}\affiliation{
         {Dipartimento di Fisica and INFM,
         Universit\`a di Roma {\em La Sapienza}, P.le A. Moro 2, 00185 Roma, Italy}
        }\affiliation{
         {INFM - CRS SMC,  Universit\`a di Roma {\em La Sapienza}, P.le A. Moro 2, 00185 Roma, Italy}
        }
\author{G.~Foffi}\affiliation{
         {Dipartimento di Fisica and INFM,
         Universit\`a di Roma {\em La Sapienza}, P.le A. Moro 2, 00185 Roma, Italy}
        }
\author{F.~Sciortino} \affiliation{
         {Dipartimento di Fisica and INFM,
         Universit\`a di Roma {\em La Sapienza}, P.le A. Moro 2, 00185 Roma, Italy}
        }\affiliation{
        {INFM - CRS Soft,  Universit\`a di Roma {\em La Sapienza}, P.le A. Moro 2, 00185 Roma, Italy}
        }
\author{P.~Tartaglia} \affiliation{
         {Dipartimento di Fisica and INFM,
         Universit\`a di Roma {\em La Sapienza}, P.le A. Moro 2, 00185 Roma, Italy}
        }\affiliation{
         {INFM - CRS SMC,  Universit\`a di Roma {\em La Sapienza}, P.le A. Moro 2, 00185 Roma, Italy}
        }


\begin{abstract}
We study tagged particle diffusion at large packing fractions, for a
model of particles interacting with a generalized Lennard-Jones
$2n$-$n$ potential, with large $n$.  The resulting short-range
potential mimics interactions in colloidal systems.  In agreement with
previous calculations for short-range potential, we observe a
diffusivity maximum as a function of temperature.
By studying the temperature dependence of the configurational entropy
--- which we evaluate with two different methods --- we show that a
configurational entropy maximum is observed at a temperature close to
that of the diffusivity maximum. Our findings suggest a relation
between dynamics and number of distinct states for short-range
potentials.
\end{abstract}
\pacs{82.70.Dd, 64.70.Pf, 61.20.-p}

\maketitle


In recent years, several studies have been focused on short-range
attractive systems, in which the range of the attractive part of the
potential is much shorter then the typical repulsive core. In nature
this behavior is encountered in colloidal systems, in which the range
of the attraction can be finely tuned by changing the solvent
properties or the chemistry of the dispersed particles. The short
length-scale of the attraction generates many unusual phenomena, (for
recent reviews see for example
Ref.~\onlinecite{Frenkel2002,Anderson2002}). From a thermodynamical
point of view, it is now well established that the liquid-liquid
coexistence becomes metastable with respect to the fluid-solid one
\cite{Gast83,Meijer1991}, a phenomenon which is not observed
in atomic or molecular systems where the interaction  is always
long-ranged.  Recently, theoretical, numerical and experimental
studies have analyzed the dynamical properties of these
systems. One of the most astonishing discoveries is that, at high
density, the metastable liquid is characterized by a non-monotonic
temperature ($T$)  dependence of the diffusivity: dynamics slows down not
only upon cooling (as is commonly observed in molecular systems), but
also upon heating. The slowing down upon heating can be so intense
that a novel mechanism of arrest takes place at high $T$
\cite{Sciortino2002b}, creating in the same system two different glass phases: one at
low $T$ called the {\it attractive glass} and one at high
$T$, the {\it repulsive glass}. The former is generated by
the short-range attractive part of the potential, while the latter is
due to the cage effect, i.e., caging of the particle in a shell of
neighbors, analogous to the hard-sphere glass.  This scenario has been
confirmed experimentally
\cite{Mallamace2000etal,Pham2002e2004,Chen2003due,Eckert2002due}
 and by numerical simulations \cite{Puertas2002due,Zaccarelli2002be}.

The existence of a fluid phase between two glass regions suggests that
the diffusion coefficient $D$ exhibits a maximum when the strength of
the attraction (or the temperature) is varied along a path of constant
density.  Such a feature has been observed in simulations of
short-range square well models. The non-monotonic behavior of $D$ has
been explained in term of Mode Coupling Theory as resulting from the
competition of two different localization lengths, one associated with
the ``hard-core'' and one with the short-range attractive bond
\cite{Dawson2001e}.  In this Letter we calculate the $T$ dependence of
the configurational entropy $S_{conf}$ --- a measure of the number of
distinct states of the system --- with the aim of providing insights
into the physical origin of the $D$ maximum \cite{nota_water}.
We use two different routes to determine $S_{conf}$: the first one,
based on a Potential Energy Landscape (PEL) investigation, requires an
estimate of the vibrational free energy of the system close to the
explored local minima of the PEL (the so-called inherent structures
(IS) \cite{stillingeradue}); the second one uses a perturbed Hamiltonian
\cite{Frenkel2001,Coluzzi1999etal} and requires the mean-square distances from
a given equilibrium configuration in the limit of vanishing
perturbation.  In both cases, the configurational entropy is
calculated as a difference between total entropy (estimated via
thermodynamic integration from the ideal gas state) and the
vibrational entropy.  In the PEL formalism, it is crucial that the
interaction potential of the system is continuous, so that IS can be
properly located via a steepest descent minimization of the potential
energy and so that vibrational frequencies can be properly calculated.
For this reason we focus on a continuous model that possesses a steep
repulsion and a short-ranged attraction, and that has been proved to
reproduce features of short-ranged attractive colloidal system
discussed above \cite{Vliegenthart1999}.
The main results of the present work are: {\it i}) the presence of a
diffusivity maximum on varying the temperature for the considered
attractive-colloid model that confirms that origin of this phenomenon
can be ascribed to the potential being short range rather then on the
particular shape chosen
\cite{Puertas2002due,Zaccarelli2002be}; {\it ii})
the existence of a maximum in the $T$ dependence of $S_{conf}$ ---
independent of the chosen method to estimate it; {\it iii}) the fact that the
temperature at which $D$ and $S_{conf}$ have a maximum nearly coincide.

\begin{figure}[tbh]
\includegraphics[width=.5\textwidth]{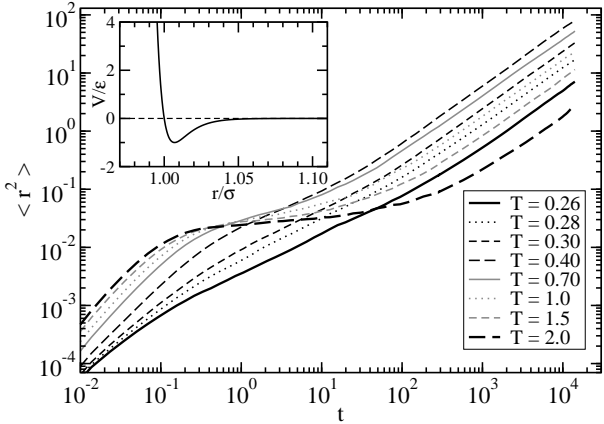}
\caption{Mean square displacements for different temperatures (from $T$=$0.26$ to $T$=$2.0$).
In the inset the pair potential LJ $2n$-$n$, with $n$=$100$.
} 
\label{fig_1}
\end{figure}

The chosen model of attractive colloid is based on a generalization
of the Lennard-Jones pair potential (LJ $2n$-$n$) proposed by
Vliegenthart {\em et al.} \cite{Vliegenthart1999}:
\begin{equation}
V_{LJn} (r) = 4 \epsilon \left[ \left( \frac{\sigma}{r} \right)^{2n} -
\left( \frac{\sigma}{r} \right)^{n} \right] \ .
\end{equation}
Differently from the original proposition we choose an extremely large
value of the exponents, i.e. $n=100$. For this value of $n$, the
potential (reported in the inset of Fig.~\ref{fig_1}) provides a
continuous version of the $3\%$ square-well potential, which has been
extensively studied in recent years~\cite{Zaccarelli2002be}.

We perform standard isothermal molecular dynamics using the
Nos\'{e}-Hoover thermostat~\cite{Frenkel2001}. The simulated system is composed of $256$
particle confined in a cubic box with periodic boundary conditions.
In order to prevent crystallization we study a $50:50$ binary mixture
 with the following parameters: $\sigma_{AA} /
\sigma_{BB}$=$1.2$, $\sigma_{AB}$=$(\sigma_{AA} +\sigma_{BB} )/ 2$,
$\epsilon_{AA}$=$\epsilon_{BB}$=$\epsilon_{AB}$. We use Lennard Jones
units ($\sigma_{AA}$ for length, $\epsilon_{AA}$ for energy,
$\tau$=$(m \sigma_{AA}^2 /\epsilon_{AA})^{1/2}$ for time). We chose
the Boltzmann constant $k_B=1$, consequently the temperature is
measured in unit of $\epsilon_{AA}$.  Cut and shift in the pair
potential are used ($r_{cut}$=$1.4$).  The time step used is
$t_0\!=\!7\!\times\!10^{-5}$.  The packing fraction investigated is
$\phi$=$0.59$, corresponding to density $\rho \simeq 1.43$.

\begin{figure}[tbh]
\includegraphics[width=.5\textwidth]{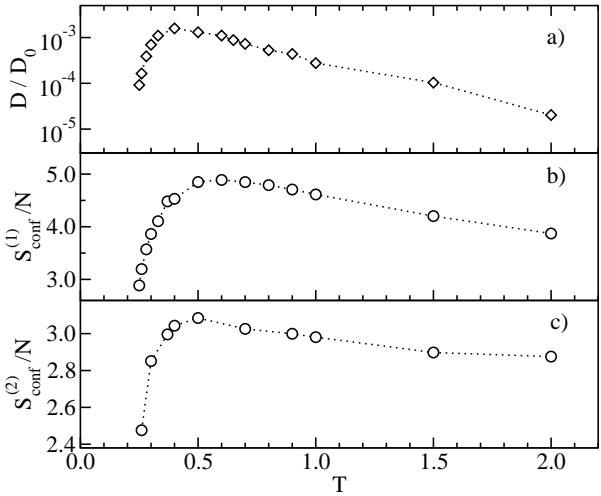}
\caption{
a) Diffusivity $D$ as a function of temperature. The value of $D$ is plotted rescaled by $D_0$=$\sqrt{\sigma^2 T /m}$.
b) Configurational entropy $S_{conf}^{(1)}$ per particle using Eq.~\ref{svib1} to estimate the vibrational
entropy ({\it PEL approach}).
c) $S_{conf}^{(2)}$ per particle using Eq.~\ref{svib2} for vibrational entropy
({\it Perturbed Hamiltonian approach}) \cite{nota_0.4}.
}
\label{fig_2}
\end{figure}

Fig.~\ref{fig_1} shows the mean-square displacement,
$r^2(t)$=$N^{-1}\langle [{\bf r}(t)-{\bf r}(0)]^2 \rangle$, where
${\bf r}$ is the $3N$ vector of the coordinates, for different
temperatures ranging from $T$=$0.26$ to $T$=$2.0$. Result are in
agreement with previous findings based on the square-well
model \cite{Zaccarelli2002be}: at high $T$ there is a well defined plateau at
about $r^2$=$2\!\times\!10^{-2}$, while at low $T$ the plateau
disappears and one observes a transient sub-diffusive regime
before the diffusive one.  It is worth noting that the value of
$r^2\!\sim\!10^{-3}$ at the beginning of the transient regime at low
$T$ corresponds to the length of the attractive well of the pair
potential.  Fig.~\ref{fig_2}a shows the $D$ (evaluated from the long
time limit behavior of $r^2$ and rescaled by the quantity
$D_0$=$\sqrt{\sigma^2 T /m}$) as a function of $T$. One observes a
maximum located at about $T_{max}=0.4$.  On lowering $T$, for
$T<T_{max}$, $D$ decreases quickly, as the system approaches the
attractive glass line (almost horizontal in the $\phi-T$ plane). On
increasing $T$, for $T>T_{max}$, $D$ decreases smoothly, consistent
with the observation that the repulsive glass line is almost vertical
in the $\phi-T$ phase diagram.


In analogy with recent studies for atomic and molecular liquids, one
can ask if the $T$ dependence of $D$ close to arrested states is
correlated to the $T$ dependence of the $S_{conf}$, as,
for example, in the well known Adam-Gibbs proposition \cite{AG}. In the
case of short-range models, a test of the correlation between $D$ and
$S_{conf}$ can be exploited in a more direct way, capitalizing on the
presence of the $D$-extremum at $T=T_{max}$.
  We try to answer this question, calculating the configurational entropy for our model system.
To estimate $S_{conf}$, we write the total entropy as a sum of two
contributions: a local vibration entropy $S_{vib}$ and a
configurational entropy $S_{conf}$ that takes into account the number
of distinct local states 
\begin{equation}
S (T,\rho) = S_{conf} (T,\rho)  + S_{vib} (T,\rho) \ .
\label{Stot}
\end{equation}
This expression, which can be formally derived within the PEL
framework~\cite{stillingeradue} and within mean field for
models of disordered $p$-spin systems \cite{Cavagna1998}, is based on the
idea that dynamics is described by two well separated time-scales: a
fast dynamics describing local rearrangements of particles within a
state (vibrations around local minima of the PEL) and a slow dynamics
which accounts for the slow exploration of different states.

Eq.~\ref{Stot} shows that $S_{conf}$ can be calculated from the
knowledge of total entropy $S$ and vibrational entropy $S_{vib}$.  The
total entropy $S$ can be calculated using thermodynamic integration
along paths in the $T$-$\rho$ plane. Without entering in details (see
Ref.~\cite{preparationetal}), we can write, measuring entropy in units of $k_B$:
\begin{equation}
S (T,\rho) = S(T_0,\rho) + \int_{T_0}^{T} \frac{dT}{T} 
\left( \frac{3}{2} N + \frac{\partial U}{\partial T} \right) \ ,
\label{S1}
\end{equation}
where $U$ is the potential energy, $T_0$ is a reference temperature
($T_0$=$0.4$ in our case), and $S(T_0,\rho)$ is given by the following
equation, which describes the path at constant $T_0$ connecting the
ideal gas to the reference state $(T_0,\rho)$,
\begin{equation}
S (T_0,\rho) = S^{id}(T_0,\rho) + \frac{U(T_0)}{T_0} - \frac{1}{T_0}
\int_{0}^{\rho} \frac{d\rho}{\rho^2}\  P_{ex} \ ,
\label{S2}
\end{equation}
where $S^{id}$ is the ideal gas contribution (which includes the
entropy of mixing, since we are dealing with a binary mixture),
and $P_{ex}$ is the excess pressure.
Performing numerical simulations at $15$ different $\rho$ values from
$0.036$ up to $1.434$, we estimated $P_{ex}(T_0,\rho)$ and $U(T_0,\rho)$,
which, together with $U(T,\rho)$ allow us to calculate with
sufficiently high precision the $T$ dependence of $S$.  The
calculation of the vibrational entropy requires some approximations.
For this reason we follow two independent different calculation
routes.  In the {\it PEL approach}, the vibrational entropy can be
calculated from the local curvature of the PEL around the explored
inherent structures \cite{stillingeradue}. In harmonic approximation,
the vibrational entropy can be written as (omitting the density
dependence):
\begin{equation}
S_{vib}^{(1)} (T) = \sum_{i=1}^{3N-3} [ 1 - \ln(\beta \hbar \omega_i ) ] \ ,
\label{svib1}
\end{equation}
where $\hbar$ is the Plank's constant, $\beta=1/ k_BT$
and $\omega_i$ are the eigenfrequencies of the Hessian matrix
evaluated at the inherent structure. The $T$ dependence, besides the
factor $\beta$, is contained in the different properties of the local
curvatures around the inherent structures explored at different
temperatures (i.e. in the $T$ dependence of the $\omega_i$
values)\cite{Lanave2003etal}.

Fig.~\ref{fig_2}b shows the configurational entropy per particle
$S_{conf}^{(1)}/N$, calculated as the difference between total entropy
in Eq.~\ref{Stot} and vibrational entropy $S_{vib}^{(1)}$ given by
Eq.~\ref{svib1}. The $T$ dependence shows a  maximum at
temperature $T_{max}^{(1)}\simeq 0.6$, slightly higher than that of
diffusivity ($T_{max}$=$0.4$). The presence of the maximum is a
remarkable result, indicating a close relationship between temperature
dependence of diffusivity and configurational entropy, even if a
quantitative coincidence of the peaks seems not to be exactly
achieved. Of course, the use of the harmonic approximation deserves
 a few remarks: while in the simple liquid the harmonic approximation
works well at sufficiently low $T$, in our case, due to the steepness
of the pair potential, the harmonic approximation is expected to break
down at a much lower $T$.  Moreover, an estimate of the anharmonic
contributions, following the techniques developed for atomic and
molecular systems is not feasible in the present case, due to the
strong $T$ dependence of the anharmonic energy.  In order to
corroborate the above finding, we calculate $S_{vib}$ using an
alternative method, based on a {\it Perturbed Hamiltonian approach}
\cite{Frenkel2001,Coluzzi1999etal,nota_coluzzi}.  One considers a perturbed
Hamiltonian:
\begin{equation}
\beta H' = \beta H + \alpha ({\bf r} - {\bf r}_0)^2 \ ,
\end{equation}
where $H$ is the original Hamiltonian, $\alpha$ is the strength of the
perturbation and ${\bf r}_0$ is a given equilibrium reference
configuration. The free energies $F(\alpha)$ of two systems with
different $\alpha$ values ($\alpha_{\infty}$ and $\alpha_0$) are related
by:
\begin{equation}
\beta F(\alpha_{\infty}) = \beta F(\alpha_0) + \int_{\alpha_0}^{\alpha_{\infty}} d\alpha' 
\langle ({\bf r} - {\bf r}_0)^2 \rangle_{\alpha'} \ ,
\label{free}
\end{equation}
where $\langle ... \rangle_{\alpha'}$ is the canonical average for a specified $\alpha'$.
In the large  
$\alpha_{\infty} \rightarrow \infty$ limit, $F(\alpha_{\infty})$ has the exact expression
\begin{equation}
\beta F(\alpha_{\infty}) = 
3 N \ln \lambda + 
\beta E_0 + \frac{3N}{2} \ln(\alpha_{\infty}/\pi) \ ,
\end{equation}
where $E_0$ is the potential energy of the reference configuration
${\bf r}_0$ and $\lambda$ is the thermal deBroglie wavelength $\lambda
= (2 \pi \beta \hbar^2 / m)^{1/2}$.  If a small $\alpha_0$ value can
be chosen in such a way that the corresponding system is equivalent to
the original system, but constrained to explore only the phase space
of one state, the estimate of $ \int_{\alpha_0}^{\alpha_{\infty}}
d\alpha'
\langle ({\bf r} - {\bf r}_0)^2 \rangle_{\alpha'}$ is sufficient to
evaluate the required vibrational free energy.

\begin{figure}[tbh]
\includegraphics[width=.5\textwidth]{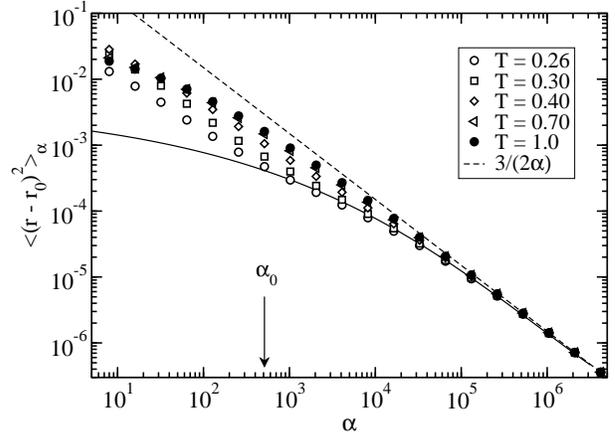}
\caption{The quantity $\langle ({\bf r} - {\bf r}_0)^2 \rangle_{\alpha}$ at different temperatures as a function 
of $\alpha$. Dashed line is the high $\alpha$ limit $3/(2\alpha)$.
Full line is a guide to the eyes of the  $T=0.26$ data, extrapolating
the approach to a {\it plateau} with a small $r^2$ value.  The arrow
indicates the value of $\alpha_0$ chosen to define a state,
$\alpha_0=5\!\times\!10^{2}$.  }
\label{fig_3}
\end{figure}

Writing the free energy as a sum of a potential energy term and an
entropic term, $\beta F(\alpha_0) = \beta E_0 + 3N/2 - S_{vib}$, the
following expression for the vibrational entropy is derived:
\begin{equation}
S_{vib}^{(2)}(T) = \int_{\alpha_0}^{\alpha_{\infty}} d\alpha' \langle ({\bf r} - {\bf r}_0)^2 \rangle_{\alpha'}
- \frac{3N}{2} \ln\left(\frac{\alpha_{\infty} \lambda^2}{\pi e}\right) \ ,
\label{svib2}
\end{equation}
where $e$ is the Neper number.  We use isothermal molecular dynamics
with Hamiltonian $H'$ to calculate $\langle ({\bf r} - {\bf r}_0)^2
\rangle_{\alpha}$ at different $\alpha$ and $T$.  We perform
averages over $20$ different reference configurations ${\bf r}_0$,
chosen from equilibrated configurations with unperturbed Hamiltonian
$H$ at temperature $T$.  Fig.~\ref{fig_3} shows $\langle ({\bf r} -
{\bf r}_0)^2 \rangle_{\alpha}$ as a function of $\alpha$ for different
$T$. The dashed line is the ($T$ independent) high
$\alpha$ limit, $3/(2 \alpha)$. As discussed above, one has to choose
an $\alpha_0$ value in Eq.~\ref{svib2} in such a way that the system
remains trapped in a given local state.  While at high temperature the
data are smooth and the values of $\langle ({\bf r} - {\bf r}_0)^2
\rangle_{\alpha}$ remain well below the cage value (about
$r^2\simeq2\!\times\!10^{-2}$), at low $T$ the behavior is quite
different: starting from high $\alpha$, we observe first an approach
to a small value $r^2\simeq2\!\times\!10^{-3}$ (corresponding to more
or less the same behavior in the mean square displacement, even if
less pronounced - see Fig.~\ref{fig_1}) and then a departure from it
at smaller values of $\alpha$.  We interpret the former as the
vibrational motion inside the state. The full line in Fig.~\ref{fig_3}
is a guide to the eyes that extrapolates the first behavior at lower
$\alpha$ for the $T$=$0.26$ case.  We have then chosen as $\alpha_0$
in Eq.~\ref{svib2} a value for which the system has not yet left the
line: $\alpha_0=5\!\times\!10^{2}$ (indicated by an arrow in
Fig.~\ref{fig_3}).  Although this is a feature only of the low $T$
data, we have chosen the same $\alpha_0$ for the estimation of
$S_{vib}^{(2)}$ for all $T$, in order to obtain a coherent definition
of it.  The $\alpha_{\infty}$ has been fixed at $2\!\times\!10^6$,
where $\langle ({\bf r} - {\bf r}_0)^2 \rangle_{\alpha}$ has reached
the asymptotic behavior (dashed line in
Fig.~\ref{fig_3}). Fig.~\ref{fig_2}c shows the $T$ dependence of the
configurational entropy per particle $S_{conf}^{(2)}/N$, calculated as
the difference between total entropy and $S_{vib}^{(2)}$ \cite{nota_0.4}.  Again one
observes a peak, located at about $T_{max}^{(2)}\simeq0.5$, close to
that of diffusivity.  We note that using the same value of $\alpha_0$
for all the temperatures introduces an underestimation of
$S_{vib}^{(2)}$, more pronounced for the high $T$ data.  This fact
could have the effect of moving the peak to a lower $T$ value,
approaching the peak value of the diffusivity.  Although the absolute
values of $S_{conf}^{(1)}$ and $S_{conf}^{(2)}$ are different, the
$T$ dependence is quite similar, suggesting that the errors are mostly
$T$ independent. More interestingly, in both approximations, a maximum
in $S_{conf}$, not far from the $T$ at which $D$ has a maximum, is
clearly detected.  Since both methods give a $T$ dependence of the
configurational entropy with a peak located close to that of
diffusivity, our work strongly support the possibility that in
short-range colloidal systems the diffusivity maximum is related to a
maximum in the number of states visited by the system.

We thank G.~Parisi for useful discussions and for suggesting the use
of the perturbed Hamiltonian method to calculate the vibrational
entropy.  We acknowledge support from INFM Initiative Parallel
Computing, and MURST COFIN2002.
\bibliography{mct,add}

\begin{thebibliography}{30}
\bibitem[{not({\natexlab{c}})}]{nota_0.4}
\bibinfo{note}{A constant $c=0.4$ has been added to the $S_{vib}^{(2)}$, to
  extimate the error originated by the finite value of $\alpha_0$ used. The
  value of $c$ has been calculated as the integral of the extrapolated line for
  $\langle(r-r_0)^2\rangle$ at $T=0.26$ (see full line in Fig.~\ref{fig_3})
  from $0$ to $\alpha_0$. This gives a correct estimation of $S_{vib}^{(2)}$
  for $T=0.26$, and represnt a lower bound for the $S_{vib}^{(2)}$ at higher
  $T$.}
\end{thebibliography}
\bibliographystyle{apsrev}

\end{document}